\documentclass[a4paper,twoside]{article}

\usepackage{epsfig}
\usepackage{subcaption}
\usepackage{calc}
\usepackage{amssymb}
\usepackage{amstext}
\usepackage{amsmath}
\usepackage{amsthm}
\usepackage{multicol}
\usepackage{pslatex}
\usepackage{apalike}
\usepackage{dirtytalk}
\usepackage{SCITEPRESS}  

\begin{document}

\title{Towards NLP-supported Semantic Data Management}

\author{\authorname{Andreas Burgdorf, André Pomp, Tobias Meisen}
\affiliation{Chair of Technologies and Management of Digital Transformation, University of Wuppertal, Wuppertal, Germany}
\email{burgdorf@uni-wuppertal.de}
}

\onecolumn \maketitle \normalsize \setcounter{footnote}{0} \vfill

\section{\uppercase{INTRODUCTION}}
\label{sec:introduction}
In the course of the progressive digitalization, the amount of data produced is growing continuously. A party that should benefit from increasing data volumes is the general public. The data stock of public institutions is growing continuously, be it by newly acquired data or by digitalized inventory data. In 2013, the G8 agreed upon the Open Data Charter with the goal to make a large part of data owned by federal institutions public so that it can be accessed by everyone and thus, promote innovation and ensure high data quality\footnote{\url{https://www.gov.uk/government/publications/open-data-charter/g8-open-data-charter-and-technical-annex}, Accessed: 2020-02-09}.

In the course of the promoted new open data strategies, multiple open data portals were launched, either operated by official side or by interest groups that engage in open data movements for certain regions or data domains. A crucial aspect for the provision of data in open data portals is metadata. These must be of high quality to ensure that data can be found and used correctly. This ensures the acceptance and increased use of these portals, especially as the Charter does not provide any proposals for standardisation and contributors are not interested in it due to its complex implementation.

This becomes even more important when data consumers like app developers are interested in data from multiple open data portals, for example, to integrate data from multiple cities, each providing its own platform. In this case, the consumer must be able to identify which data contains the same semantic information across portals.

Data providers are responsible for providing and maintaining metadata and thus, also for ensuring quality. Open data portals and their usage have been subject to many studies. In 2017, Schauppenlehner and Muhar investigated the European- and the Austrian Data Portal regarding the use of metadata \cite{schauppenlehner2018theoretical}. The authors conclude that the mere existence of metadata provides no added value. They identify weaknesses in the large number of different used data formats and in the way free text descriptions are used. Within these free texts, human-readable descriptions are mixed with machine-readable information, such as database identifiers. As consequence, data that were originally created for the human user, can only be understood by technical experts or require a high degree of pre-processing. In their review from 2012, Zuiderwijk et al. have identified in particular the characteristics of non-machine-readable metadata as a problem of open data portals \cite{zuiderwijk2012potential}. In 2016, Tygel et al. \cite{tygel2016towards} did another investigation of open data portals. They identified the usage of synonyms, ambiguity and incoherence in metadata as a key factor that hinders the reuse of open data, especially across different open data platforms. To solve this problem, they propose the integration of a semantic layer on top of multiple data platforms to improve quality and to allow interlinking multiple platforms. Their proposed approach is based on a tag manager that allows users to register, for example, synonyms when adding a data source. A prominent approach initiated by Google, Microsoft, Yahoo and Yandex is \emph{schema.org}\footnote{\url{https://schema.org}, Accessed: 2020-03-11} that has the goal to unify schemas of structured data on the web.

All in all, the use of metadata still poses challenges within individual portals and especially across several portals. Metadata are important for users to understand the semantic meaning of data. This semantic meaning should at best be comprehensible to both the computer and the human. Approaches to this are offered, for example, by ontology-based data management, which is based on the fact that the meaning of data is stored in firmly defined conceptualizations. These concepts can either be defined apriori, so that new data has to fit to the existing concepts, or the concepts can be extended step by step on the basis of new data as for example latest research by Pomp et al. \cite{pomp2019you} shows. To represent the semantic meaning of data sources, it is necessary to establish a mapping between ontology and data. This process is typically either data-driven by looking at actual values in the data, or label-driven, based on data identifiers like attribute names. Ultimately, however, human intervention is always required to define whether the concepts recognized by the algorithms are actually meaningful. Behind this step of mapping data to ontologies, there is always a certain effort, which should be kept as low as possible for the data provider so that it continues its work as continuously as possible and provides many data sources.

Especially in the case of open data portals, there is often additional data available that can support the mapping of data and ontologies and that can be created by data providers with limited effort and more informally: textual descriptions of the data in human and/or machine readable form. Considering the current research progress regarding unsupervised Natural Language Processing, new potentials arise to use these texts to extract concepts automatically. During the last two years models outperforming each other in NLP tasks like language understanding are released in short time windows. ELMo \cite{elmo}, Bert \cite{bert} and XLNnet \cite{xlnet} are some of the latest developments in this field and achieve scores of up to 90\% for reading comprehension tasks. The use of modern NLP models brings opportunities in many cases where classical statistical NLP methods were not sufficient. It is therefore quite conceivable that the use of modern NLP methods can be used to further support people in the preparation of semantic models and thus improve the harmonization of metadata.

\section{\uppercase{RESEARCH PROBLEM}}
\label{sec:problem}
As described at the beginning, using the example of open data portals, the use of metadata alone is not sufficient to ensure uniform usability and quality of the data provided and that provision of metadata is always a time-consuming task for data providers. Therefore, this task of metadata provision is carried out in different ways and with varying quality. As a result, it is particularly difficult to identify and share semantically similar data. Ontologies are one possibility for better semantic description of data, but it is necessary to develop corresponding concepts in advance or when the data to be described is made available. In this case, the expert knowledge and time expenditure of the data provider are necessary to achieve optimal results. Recent automatic approaches for semantic labeling and semantic modeling, meaning the creation of a mapping between data and ontology, are based either on the data identifiers or on the actual data values. Metadata in form of free text, as it is for example often used in open data portals, represents a third possible source for this modelling if they contain a detailed description of the dataset. With the help of natural language processing methods, among others, these texts could be used to automatically suggest concepts behind provided datasets, so that recognized concepts can subsequently be edited or released by the provider. Especially the mixed use of human and computer-readable information, which has been criticized in previous studies, represents a potential here. A hybrid approach combining existing texts and the expertise of the provider could complement the existing procedures for semantic labeling that are used to manage data semantically, using ontologies. Thus, data from different sources could be integrated into a common database in the future without the providing person having to describe the underlying concept for each dataset completely formally. An additional problem that arises from this is the handling of concepts that are not yet known in the existing ontology. This ontology should continuously evolve based on texts and user input. To formalize the problem, we can make the following assumptions: $D_i$ is an arbitrary structured dataset, $S_{D_i}$ is the actual structure of the dataset, $T_{D_i}$ is the textual description of the dataset and $M_{D_i}$ is its semantic model. $O$ is the global ontology which is empty at the beginning. We want to find out, if it is possible, to

\begin{itemize}
    \item use the description $T_{D_i}$ in combination with the ontology $O$ to receive the semantic model $M_{D_i}$ that is semantically equivalent to $S_{D_i}$
    \item use the description $T_{D_i}$ and the model $M_{D_i}$ to evolve the ontology $O$
    \item use the ontology $O$ and the semantic model $M_{D_i}$ to generate a textual description $T_{D_i}$.
\end{itemize}

\section{\uppercase{OUTLINE OF OBJECTIVES}}
\label{sec:objectives}
\begin{figure*}
    \centering
    \includegraphics[width=0.99\textwidth]{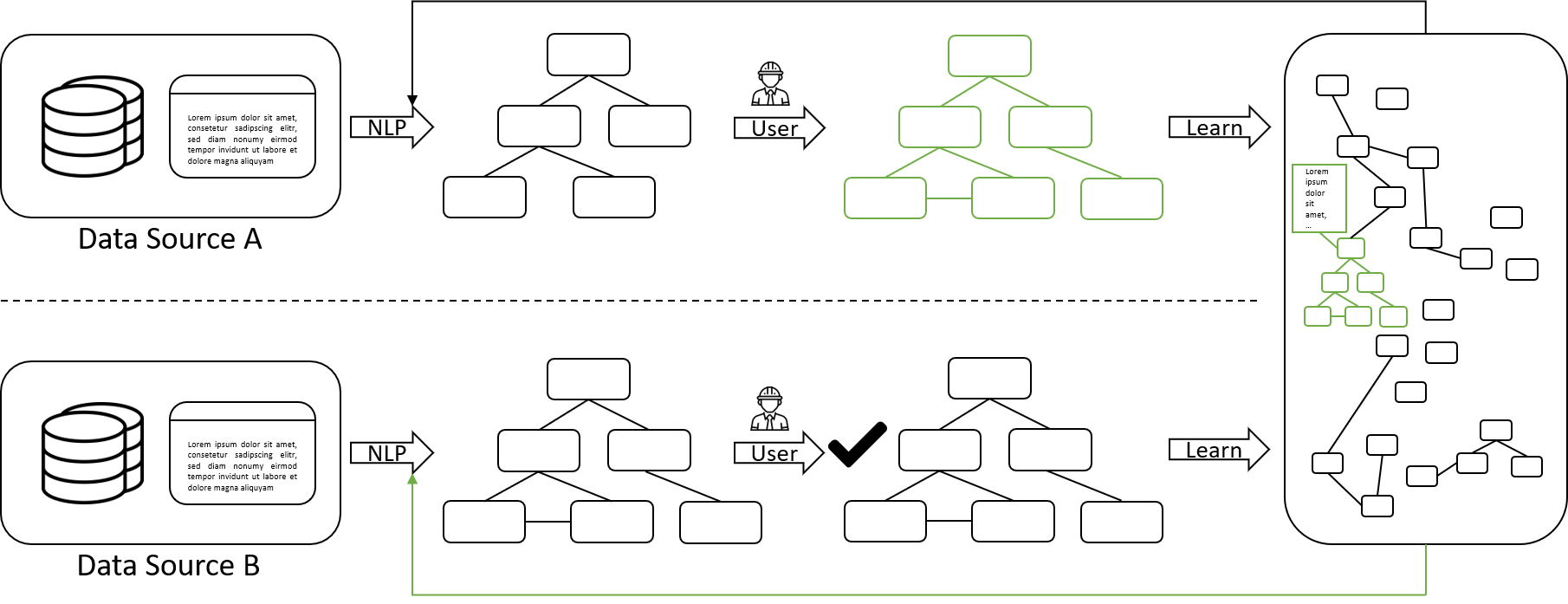}
    \caption{This figure shows our concept of adding two semantically identical data sources. Before adding any of the two data sources, the ontology on the right only consists of the black concepts. When data source A is integrated, our NLP-based approach extracts a semantic model with support of the ontology. The automatically extracted model has to be modified by the user, as it is not completely correct. The corrected version of the model is used to enhance the ontology. New concepts and their textual descriptions are added. When data source B is integrated, those new green concepts are already part of the ontology. The automatically generated semantic model is correct and the user just has to improve it instead of modifying it.}
    \label{fig:concept1}
\end{figure*}

With our research we will extend the possibilities of semantic modeling, available so far, to get a global view at the meaning behind datasets from multiple sources and multiple providers. We will leverage as many available metadata to keep the human effort for semantic modeling as small as possible and nevertheless allow to have a common semantic model across multiple datasets later on. Figure \ref{fig:concept1} visualizes how we imagine a solution here. We focus on two key issues: first, semantic modeling itself, i.e. the process of extracting concepts from a dataset consisting of the actual data and a corresponding textual description, and second, how we can use these extracted concepts to build a global ontology that can improve this process in the future.
As input data we use textual descriptions of datasets and some global ontology describing data characteristics. This global ontology, which does not contain any concepts at the beginning, will grow by adding new data sources and their semantic modeling. Texts can follow different patterns by either describing the datasets completely informal or by also integrating certain available data identifiers. We will use state of the art NLP methods to process the textual descriptions in order to identify underlying semantic concepts. This process of mapping to a semantic concept should be supported by the global ontology as well as by potentially available external data sources like domain specific thesauri or knowledge bases. After our algorithms have generated a semantic model, the data provider has to check the resulting model. He can make modifications and finally release the model if it meets his expectations. Afterwards we use the semantic model, verified by this hybrid approach to improve the global ontology in the case that previously unknown concepts were created. Furthermore, we would like to give data providers the opportunity to have descriptions of textual metadata evaluated. An evaluation here means that the description is compared to semantically similar datasets to determine if certain standards are available and if the description is sufficient to map it to a similar semantic concepts. When developing ontologies, we look beyond existing approaches to the possible use of embeddings. We will investigate the added value of storing extracted concepts and associated texts as embeddings in the ontology instead of just the confirmed concepts in graph form. Current research on text and graph embeddings promises a benefit that should be investigated in any case. The storage of embeddings could also lead to the automatic generation of descriptive texts for data sources that are entered without descriptive texts, based on the data- or label-driven semantic modelling.

In summary, our goals lead to the following research questions which we would like to answer:
\begin{enumerate}
    \item What information has to be included in textual metadata and how informal is the description allowed to be for the creation of semantic models?
    \item What are suitable methods to generate a semantic model for a data source using informal texts and existing formal concepts like ontologies?
    \item Which methods are suitable for extending an already existing ontology based on extracted semantic models?
    \item To what extent is it possible, on the basis of learned concepts and texts, to generate textual metadata for datasets whose semantic models were created without the use of texts?
\end{enumerate}

\section{\uppercase{STATE OF THE ART}}
\label{sec:sota}

\subsection{Ontologies}
In order to describe the content of data in a structured way it is necessary to formalize the existing knowledge about the dataset and the domain data comes from. A popular approach for formalizing this knowledge are ontologies. Studer et al. \cite{studer1998knowledge} define ontologies as an \say{explicit, formal specification of a shared conceptualization}. To make it more tangible, ontologies can be described as a set of concepts, which are in a certain relation to each other. For example the concept \texttt{GEO\_POSITION} could be related to the concepts \texttt{LATITUDE} and \texttt{LONGITUDE} with the relation \texttt{HAS\_ATTRIBUTE}. Ontologies are an enabler for further developments like the Semantic Web, providing a deeper understanding to resources from the web. People, responsible for building the ontology are so called ontology engineers \cite{maedche2001ontology}. A research discipline, dealing with supporting ontology engineers during creation by automating this process as far as possible, is \emph{ontology learning}. This is an important task to take unnecessary workload away from engineers or at least to support them in identifying needed concepts.
\subsection{Ontology Learning}
\label{subsec:ontologylearning}
While in classical approaches, ontology engineers are the persons, responsible for building up an ontology manually, examples from the real world show that such ontologies can grow immeasurably which leads to a high creation and correction effort. When domains are added or extended, this leads to a manual maintenance of the ontology later on \cite{maedche2001ontology}. For this reason, methods for automated ontology generation and maintenance are in the focus of research for more than 20 years now, as Ding and Foo (2002) already provided a survey on this research topic in 2002. One domain for automated generation deals with ontology learning from text. Wong et al. define this as \say{process of identifying terms, concepts, relations, and optionally, axioms from textual information and using them to construct and maintain an ontology} \cite{wong2012ontology}. As there are multiple surveys from 2019, presenting existing approaches for ontology learning from texts, we just want to give a brief overview. In a survey from 2019, Lourdusamy and Abraham \cite{lourdusamy2019survey} provide an overview of methods and frameworks for ontology learning from text. They present multiple approaches, their input and the technique they use. The input is (semi-) structured or unstructured text, as well as external knowledge in form of already existing ontologies. Presented methods mainly make use of statistical methods, Natural Language Processing (NLP) and Machine Learning (ML). Aside from the used methods it is noticeable that most approaches make use of either external taxonomies like WordNet or already existing ontologies for the specific domain to improve or extract the ontology from text. During their survey, they identified several gaps in the reviewed approaches. Beyond others these gaps are not satisfying methods for language understanding and knowledge extraction and the problem of dealing with noise in text data. Lourdusamy and Abraham (2019) come to the conclusion that by now the fully automated ontology creation from text is not possible, but semi-automated approaches are. Another method is presented by Xu et al. \cite{xu2019automatic} in 2019. They present the creation of ontologies from unstructured text by using a two classifier approach. In a first step it identifies candidates for concepts in the text, in the second step it assigns them to actual concepts. Their approach is based on linguistic features, the context of words inside a sentence and they make use of pre-trained word embeddings.

\subsection{Ontology-Based Data Management}
In 2011, Lenzerini \cite{lenzerini2011ontology} defined the paradigm of ontology-based data management (OBDM). He defines the goals as unified access to data as well as the governance of processes for integrating and handling data. The necessity for this paradigm arose from practical problems of data governance as experienced by companies in the context of growing data volumes. The author describes that the fact that database structures in companies change over time and are adapted to the application using them creates a kind of silo structure: there are a number of databases that can each be used by one application, but which are otherwise independent of each other and that hardly enable any unified data access. The idea of OBDM is establishing a three-level architecture for data management:

\paragraph{Ontology Layer:} The ontology layer represents the formal description of the domain, that the data comes from or is used for.
\paragraph{Source Layer:} The source layer represents the actual data in a structure that is independent from any application specific requirements.
\paragraph{Mapping Layer:} The mapping layer represents the relations between concepts from the ontology layer and the corresponding data sources from the source layer.

The author mentions as an advantage of this procedure that the knowledge from the ontologies, unlike with pure database structures, is reusable, that an explicit documentation of data sources is available through the mapping layer and that data sources do not have to be completely integrated immediately, but that the system can evolve over time \cite{lenzerini2011ontology}.

The OBDM paradigm can be found today in several software solutions. Examples are \emph{KARMA} \cite{knoblock2012semi}, \emph{Optique} \cite{kharlamov2013optique} or \emph{ESKAPE} \cite{pomp2017eskape}. KARMA is located at the Semantic Web, allowing users to integrate data from multiple sources. The process is supported by a graphical user interface. Users model their data with respect to an ontology and the system creates the mappings between data and ontology. Optique is developed close to industry and optimized for accessing data from big data sources efficiently by using an ontology, mapped to the data sources. ESKAPE allows users to integrate data sources by adding a semantic model for the data. Based on the providing models, a knowledge-graph is evolving and supports later data integration and information-driven access to data.

\subsection{Semantic Labeling and Modelling}
When implementing the mapping layer, \emph{semantic labeling} or \emph{semantic modelling} comes to the fore. While semantic labeling creates pairs of attributes from the data source and the specific attributes from an ontology, they belong to, semantic modelling also takes relations between attributes into account \cite{vu2019learning}. Instead of building a map, semantic modelling builds a tree in which the nodes define attributes and the edges define relationships. Since there is a danger of confusion between similarly named attributes, the automated creation of semantic models is still a great challenge, which is why the user is usually supported but not replaced by automated systems. Vu et al. \cite{vu2019learning} classify existing approaches for automated semantic modelling into those that try to match data schemas and those that describe the semantic using external knowledge like DBpedia for finding relationships. The latter one can be driven by attribute names as well as by data values.

\subsection{Extracting Models from Text}
As, to the best of our knowledge, there currently is little research on the usage of text for semantic modelling, we want to give a brief overview of research that might be relevant here.  Beyond the here presented approaches, methods for ontology learning from text, as discussed by Wong et al. \cite{wong2012ontology} could be taken into account, here as well. There is many domain specific research available that focuses on extracting certain use case-relevant information from texts. Wang et al. \cite{wang2016using} present a more generic approach that extracts knowledge graphs from texts. With their approach they tackle the problem of \emph{key concept extraction} and \emph{concept relationship identification} being both extracted in form of the knowledge graph. For the extraction they make use of word embeddings, word based similarities and external sources like Wikipedia. Using knowledge graphs here, this research is already closely related to our problem of extracting semantic models. Si et al. \cite{si2019enhancing} demonstrate the possibilities of modern global NLP models like ELMo \cite{elmo} or Bert \cite{bert} for the task of concept extraction with the example of medical documents. By using global contextualized embeddings they achieve almost as good results as by using highly domain-specific embeddings.

\section{\uppercase{METHODOLOGY}}
\label{sec:methodology}
From the problem we described and the first look at state of the art, we can define a first plan on how to tackle the problem during the next years in the scope of further scientific publications.

\paragraph{Phase 1: Identification of Methods and Datasets}
The goal of the first phase is to identify methods and data to work with for the further course. Following our initial literature research, we are going to examine, which methods are suitable best for deriving metadata concepts from textual descriptions and to use them for semantic modeling in combination with human input. We want to analyze similarities and differences in existing methods, characterize advantages and disadvantages of each and prepare the results in a survey paper on methods. By now, there is no existing dataset which will help us in developing and evaluating our own methods. To build our own evaluation dataset, we will do a research across multiple relevant open data portals to find out, which kind of textual metadata is used. Based on the findings we will setup a suitable evaluation dataset, consisting of structured data in combination with textual metadata.

\paragraph{Phase 2: Method Development and Metric Definition}
As second step, we will examine, how the identified methods can be combined and how they have to be extended in order to realize the proposed hybrid approach. In the course of this, we will develop an actual solution that leverages existing metadata to support users with the task of annotating their datasets. Besides the development of methods, a user study to evaluate the benefit for users in terms of time saving will be necessary. For this, it is important to identify suitable metrics at this stage: how can the benefit of users be measured? Possible dimensions might be saving time or increasing quality or standardization.

\paragraph{Phase 3: Metadata Assessment}
Next, we will find out, how we can measure the usefulness of newly provided metadata, based on the continuously growing global ontology. When users provide a new dataset they are in the duty to provide sufficient metadata as well if they want to keep the effort of manual semantic modelling small. By comparing the new metadata to already known historical data and the ontology, it is likely that we can predict beforehand, how much manual effort data providers have to invest after integration. Thus, we are looking for some kind of forecast to predict the degree of manual modifications of the semantic model, needed later on. In this context, the evaluation dataset, created in context of phase 1, has to be extended and manually annotated to allow for meaningful evaluation results, regarding the similarity between different datasets, regarding their content and their semantic modeling effort.

\paragraph{Phase 4: Turning Things Around}
Finally, we assume that the creation of semantic models of data has been successful. Then, these models are helpful for the computer for better understanding of the data structure but not for the average human user who wants to access datasets. To create another benefit for human users, we will find out if it is possible to leverage the created semantic models and the ontology to automatically generate a textual description of a new dataset. Instead of using textual descriptions for semantic modeling, only a data- or label-driven method will be used for semantic modelling in this phase. Finding concepts of semantically similar data sources and their textual description in the global ontology, we will evaluate if the identified concepts and information from similar earlier concepts are sufficient to generate a human-readable description text for the dataset, automatically.

\section{\uppercase{EXPECTED OUTCOME}}
\label{sec:outcome}
As soon as we are able to obtain relevant findings from our research, we will publish them. Developed methods and software, as well as built datasets we would like to make available to the community for discussion in the course of further scientific publications. We will orient ourselves along the phases described above:

\paragraph{Outcome Phase 1:}
In the first phase, we will publish a review of methods suitable for the extraction of a semantic model from textual metadata at a very early stage, as it is fundamental for all subsequent work. Further, we will create an evaluation dataset that contains a large number of datasets among which semantic similarities exist. For each of these datasets, our evaluation dataset will also contain a textual description. Due to high availability, we will compile the data from a variety of open data portals. The datasets will be from different domains (e.g. geodata and economic data), from different regions and especially from different data providers. This should ensure a heterogeneity in the metadata with which we can ensure that our approach is not only working domain-specifically.
\paragraph{Outcome Phase 2:}
As during the second phase, method development and metric definition are in the focus, this will be reflected in the outcome. At first we will present the methods we develop as a concept in a publication. In a second publication we will present the actual implementation of the method, a suitable metric to evaluate the method and the results that we achieve on the evaluation dataset that resulted from the first phase. We will to make the actual software implementation available as open source, so that methods and achieved results are comprehensible for everyone interested.
\paragraph{Outcome Phase 3:}
As mentioned above, in course of phase three, the evaluation dataset has to be extended. We will publish the extended dataset, also containing the semantic modeling effort score, to check if unknown data can be assessed correctly. In a scientific publication we will publish our method for data assessment and the results that we can achieve on the provided dataset. Resulting implementations will be made available as open source tools.
\paragraph{Outcome Phase 4:}
Since we would like to find out in the fourth phase if automated text generation based on historical data is possible, we will publish a paper that gives an overview of existing methods for automated text generation. Based on this, we will present a suitable methodology for our application case in a scientific publication. To what extent a further data set can be created here is still open, but it should be possible to use the data set created in phase 1 for a manual comparison of the generated texts with real texts. An automatic comparison of generated texts with manually provided by humans will be hard to realize, as the goal of our method is the human text understanding.

\section{\uppercase{STAGE OF THE RESEARCH}}
\label{sec:stage}
The research for the thesis is still at a very early stage. At this point we have gained a first insight into the different existing Open Data Portals and have looked at the different ways metadata is handled. In addition, we have started a first analysis of the current state of research, which has to be intensified in order to identify current lines of research that may not be related to the project at first sight, but which can provide further insights.

\section{\uppercase{CONCLUSION}}
\label{sec:conclusion}
In this paper, we presented the problem of generating semantic models for managing shared data, using the model to enhance ontologies and using ontologies to create textual descriptions. With the example of open data portals we showed that metadata alone is not sufficient to provide all semantic relevant information of data. While the research field of ontology-based data management already deals with the two dimensions data labels and data values for the automatic generation of semantic models from data, we identified a third dimension that might be useful: textual descriptions. Using modern NLP approaches, an automated extraction of concepts from texts is promising. 
With the identified problem we have set our own research goals. We will use the textual descriptions of data to extract semantic models that can be modified or approved by human users. Based on approved models we will develop and extend a global ontology, that can be used to improve the process of automatic semantic model creation, but also offers the possibilities to assess unknown datasets or to create textual descriptions automatically. Based on these questions we provided a brief overview of the state of the art regarding ontology-based data management, ontology learning and semantic mapping. Finally, we divided the necessary research work into four phases and showed how we are going to proceed in the current phases and what results we expect. In particular, methods, large evaluation datasets and implementations will be published.

\bibliographystyle{apalike}
{\small
\bibliography{main}}

\begin{thebibliography}{}

\bibitem[Devlin et~al., 2018]{bert}
Devlin, J., Chang, M.-W., Lee, K., and Toutanova, K. (2018).
\newblock Bert: Pre-training of deep bidirectional transformers for language
  understanding.
\newblock {\em arXiv preprint arXiv:1810.04805}.

\bibitem[Kharlamov et~al., 2013]{kharlamov2013optique}
Kharlamov, E., Giese, M., Jim{\'e}nez-Ruiz, E., Skj{\ae}veland, M.~G., Soylu,
  A., Zheleznyakov, D., Bagosi, T., Console, M., Haase, P., Horrocks, I.,
  et~al. (2013).
\newblock Optique 1.0: semantic access to big data: the case of norwegian
  petroleum directorate's factpages.
\newblock In {\em Proceedings of the 12th International Semantic Web Conference
  (Posters \& Demonstrations Track)-Volume 1035}, pages 65--68. CEUR-WS. org.

\bibitem[Knoblock et~al., 2012]{knoblock2012semi}
Knoblock, C.~A., Szekely, P., Ambite, J.~L., Goel, A., Gupta, S., Lerman, K.,
  Muslea, M., Taheriyan, M., and Mallick, P. (2012).
\newblock Semi-automatically mapping structured sources into the semantic web.
\newblock In {\em Extended Semantic Web Conference}, pages 375--390. Springer.

\bibitem[Lenzerini, 2011]{lenzerini2011ontology}
Lenzerini, M. (2011).
\newblock Ontology-based data management.
\newblock In {\em Proceedings of the 20th ACM international conference on
  Information and knowledge management}, pages 5--6.

\bibitem[Lourdusamy and Abraham, 2019]{lourdusamy2019survey}
Lourdusamy, R. and Abraham, S. (2019).
\newblock A survey on methods of ontology learning from text.
\newblock In {\em International Conference on Information, Communication and
  Computing Technology}, pages 113--123. Springer.

\bibitem[Maedche and Staab, 2001]{maedche2001ontology}
Maedche, A. and Staab, S. (2001).
\newblock Ontology learning for the semantic web.
\newblock {\em IEEE Intelligent systems}, 16(2):72--79.

\bibitem[Peters et~al., 2018]{elmo}
Peters, M.~E., Neumann, M., Iyyer, M., Gardner, M., Clark, C., Lee, K., and
  Zettlemoyer, L. (2018).
\newblock Deep contextualized word representations.
\newblock {\em arXiv preprint arXiv:1802.05365}.

\bibitem[Pomp et~al., 2019]{pomp2019you}
Pomp, A., Lipp, J., and Meisen, T. (2019).
\newblock You are missing a concept! enhancing ontology-based data access with
  evolving ontologies.
\newblock In {\em 2019 IEEE 13th International Conference on Semantic Computing
  (ICSC)}, pages 98--105. IEEE.

\bibitem[Pomp et~al., 2017]{pomp2017eskape}
Pomp, A., Paulus, A., Jeschke, S., and Meisen, T. (2017).
\newblock Eskape: Information platform for enabling semantic data processing.
\newblock In {\em ICEIS (2)}, pages 644--655.

\bibitem[Schauppenlehner and Muhar, 2018]{schauppenlehner2018theoretical}
Schauppenlehner, T. and Muhar, A. (2018).
\newblock Theoretical availability versus practical accessibility: The critical
  role of metadata management in open data portals.
\newblock {\em Sustainability}, 10(2):545.

\bibitem[Si et~al., 2019]{si2019enhancing}
Si, Y., Wang, J., Xu, H., and Roberts, K. (2019).
\newblock Enhancing clinical concept extraction with contextual embeddings.
\newblock {\em Journal of the American Medical Informatics Association},
  26(11):1297--1304.

\bibitem[Studer et~al., 1998]{studer1998knowledge}
Studer, R., Benjamins, V.~R., and Fensel, D. (1998).
\newblock Knowledge engineering: principles and methods.
\newblock {\em Data \& knowledge engineering}, 25(1-2):161--197.

\bibitem[Tygel et~al., 2016]{tygel2016towards}
Tygel, A., Auer, S., Debattista, J., Orlandi, F., and Campos, M. L.~M. (2016).
\newblock Towards cleaning-up open data portals: A metadata reconciliation
  approach.
\newblock In {\em 2016 IEEE Tenth International Conference on Semantic
  Computing (ICSC)}, pages 71--78. IEEE.

\bibitem[Vu et~al., 2019]{vu2019learning}
Vu, B., Knoblock, C., and Pujara, J. (2019).
\newblock Learning semantic models of data sources using probabilistic
  graphical models.
\newblock In {\em The World Wide Web Conference}, pages 1944--1953.

\bibitem[Wang et~al., 2016]{wang2016using}
Wang, S., Ororbia, A., Wu, Z., Williams, K., Liang, C., Pursel, B., and Giles,
  C.~L. (2016).
\newblock Using prerequisites to extract concept maps fromtextbooks.
\newblock In {\em Proceedings of the 25th acm international on conference on
  information and knowledge management}, pages 317--326.

\bibitem[Wong et~al., 2012]{wong2012ontology}
Wong, W., Liu, W., and Bennamoun, M. (2012).
\newblock Ontology learning from text: A look back and into the future.
\newblock {\em ACM Computing Surveys (CSUR)}, 44(4):1--36.

\bibitem[Xu et~al., 2019]{xu2019automatic}
Xu, Y., Rajpathak, D., Gibbs, I., and Klabjan, D. (2019).
\newblock Automatic ontology learning from domain-specific short unstructured
  text data.
\newblock {\em arXiv preprint arXiv:1903.04360}.

\bibitem[Yang et~al., 2019]{xlnet}
Yang, Z., Dai, Z., Yang, Y., Carbonell, J., Salakhutdinov, R.~R., and Le, Q.~V.
  (2019).
\newblock Xlnet: Generalized autoregressive pretraining for language
  understanding.
\newblock In {\em Advances in neural information processing systems}, pages
  5754--5764.

\bibitem[Zuiderwijk et~al., 2012]{zuiderwijk2012potential}
Zuiderwijk, A., Jeffery, K., and Janssen, M. (2012).
\newblock The potential of metadata for linked open data and its value for
  users and publishers.
\newblock {\em JeDEM-eJournal of eDemocracy and Open Government},
  4(2):222--244.

\end{thebibliography}

\end{document}